\documentclass{article}

\usepackage{PRIMEarxiv}

\usepackage[utf8]{inputenc} 
\usepackage[T1]{fontenc}    

\usepackage{url}            
\usepackage{booktabs}       
\usepackage{amsfonts}       
\usepackage{nicefrac}       
\usepackage{microtype}      
\usepackage{lipsum}
\usepackage{fancyhdr}       
\usepackage{amsmath}
\usepackage{moreverb}

\usepackage[colorlinks,bookmarksopen,bookmarksnumbered,citecolor=red,urlcolor=red]{hyperref}

\usepackage{graphicx}

\title{Theoretical scheme on shape-programming of thin hyperelastic plates through differential growth}

\author{
  Jiong Wang$^{1,2}$,  Zhanfeng Li$^1$,Zili Jin$^1$\\
  $^1$School of Civil Engineering and Transportation, South China University of Technology, China \\
  $^2$State Key Laboratory of Subtropical Building Science, South China University of Technology, China \\
  Correspondent author: Jiong Wang, School of Civil Engineering and Transportation, \\
  South China University of Technology, 510640 Guangzhou, Guangdong, China  \\
  Email: ctjwang@scut.edu.cn \\
}

\begin{document}
\maketitle
\rhead{Wang et al.}
\lhead{ }

\begin{abstract}

In this paper, a theoretical scheme is proposed for shape-programming of thin hyperelastic plates through differential growth. First, starting from the 3D governing system of a hyperelastic (neo-Hookean) plate, a consistent finite-strain plate equation system is formulated through a series-expansion and truncation approach. Based on the plate equation system, the problem of shape-programming is studied under the stress-free assumption. By equating the stress components in the plate equations to be zero, the explicit relations between growth functions and geometrical quantities of the target shape of the plate are derived. Then, a theoretical scheme of shape-programming is proposed, which can be used to identify the growth fields corresponding to arbitrary 3D shapes of the plate. To demonstrate the efficiency of the scheme, some typical examples are studied. The predicted growth functions in these examples are adopted in the numerical simulations, from which the target shapes of the plate can be recovered completely. The scheme of shape-programming proposed in the current work is applicable for manufacture of intelligent soft devices.

\end{abstract}


\maketitle


\section{Introduction}
\label{sec:1}

Growth (or swelling) of soft material samples (e.g., soft biological tissues, polymeric gels) are commonly observed in nature and in engineering fields \cite{ambr2011,liu2015,ambr2019}. Due to the factors of genetic, biochemistry, environmental stimuli and mechanical loads, the growth fields in soft material samples are usually inhomogeneous or incompatibility, which is referred to as differential growth. In this case, the soft material samples can exhibit diverse geometrical shape changes and surface pattern evolutions during the growing processes \cite{holm2011,li2012,kemp2014,huan2018}. On the other hand, through elaborate design of the compositions or architectures in the soft material samples, it is possible to control the growth-induced deformations of the samples such that certain intended configurations are fabricated or other kinds of functions are realized. This procedure is known as `shape-programming' \cite{liu2016} and it has been utilized for the manufacture of novel intelligent soft devices, e.g., actuators, sensors, soft robotics \cite{iono2013,glad2016,yuk2017,sief2019}.

Within the range of nonlinear elasticity, soft materials can be viewed as hyperelastic materials \cite{ogden1984}. To take the growth effect into account, the total deformation gradient tensor is usually decomposed into the multiplication of an elastic deformation tensor and a growth tensor \cite{kond1987,rodr1994,amar2005}. In another modeling approach proposed based on the geometric theory, the growth effect is interpreted as the metric evolution in the material manifolds \cite{efra2009,yava2010}. Based on the materials' constitutive assumptions and through some conventional approaches, the governing equation system for modeling the growth-induced deformations of soft material samples can be established. As the elastic deformations of soft materials are generally isochoric, the constraint equation of elastic incompressibility should also be adopted. Most of the existing modeling works focus on the direct problem. That is, by specifying the given growth fields (or growth functions) in the soft material samples, which kind of mechanical behaviors will be exhibited by the samples. The results of these works can usually provide good simulations on the growing processes of soft biological tissues in nature \cite{gori2017,liu2020,xu2020,chen2021}. While, to fulfill the requirements of shape-programming, one also needs to consider an inverse problem. That is, to achieve certain target shapes through differential growth, how to arrange the growth fields in the soft material samples?

Regarding the inverse problem, some research works have been reported in the literature. In these works, the initial configurations of soft material samples usually have the thin plate form. Dias et al. \cite{dias2011} studied the generations of particular three-dimensional (3D) shapes from thin elastic sheets by mere imposition of a two-dimensional (2D) pattern of locally isotropic growth, where the non-Euclidean plate model proposed in Efrati et al.\cite{efra2009} was adopted. Jones and Mahadevan \cite{jone2015} proposed a numerical approach to determine the optimal growth field giving rise to an arbitrary target shape of the soft material sample. Wang et al. \cite{wang2019} focused on the plane-strain problems and derived some explicit analytical formulas for 2D shape-programming of hyperelastic plates through differential growth. Nojoomi et al. \cite{nojo2021} designed the scheme of 2D growth for target 3D shapes via conformal flattening and incorporated the concept of cone singularities to increase the accessible space of 3D shapes. Despite the existences of these works, the current research state on the inverse problem has not attained a satisfactory level. To our knowledge, the existing analytical results for this inverse problem is rare and there is still no theoretical scheme that can be used to generate arbitrary 3D target shapes of soft material samples.

In the current work, we aim to propose a theoretical scheme for shape-programming of thin hyperelastic plates through differential growth. The basis of the current work is a consistent finite-strain plate theory proposed in Wang et al. \cite{wang2018}. The plate equation system in this theory is derived from the 3D governing system through a series expansion and truncation approach \cite{Dai2014}, which incorporates the growth effect and the constraint of elastic incompressibility. With the established plate equation system, we focus on the stress-free condition. By equating the stress components in the plate equations to be zero, the explicit relations between growth functions and geometrical quantities of the target shapes of the plate are derived, which have some relatively simple forms. By virtue of these relations, a theoretical scheme of shape-programming is proposed, which can be used to identify the growth fields corresponding to arbitrary 3D target shapes of the plate. To demonstrate the correctness and efficiency of the scheme, some typical examples are studied. In these examples, the growth functions are determined by using the theoretical scheme, which are further adopted in the numerical simulations. It will be seen that with predicted growth functions, the target shapes of the plate can be recovered completely in the numerical simulation results.

This paper is organized as follows. In Section 2, the plate equation system for modeling the growth-induced deformation of a thin hyperleastic plate is formulated. In section 3, the analytical relations between growth functions and geometrical quantities of the target shapes of the plate is derived, based on which a theoretical scheme for shape-programming of hyperelastic plates is proposed. In section 4, some typical examples are studied to demonstrate the correctness and efficiency of the scheme. Finally, some conclusions are drawn.


\section{Finite-strain plate theory with growth effect}
\label{sec:2}


\subsection{Preliminaries}
\label{sec:2.1}

Consider a thin hyperelastic plate with constant thickness, which locates in the three-dimensional (3D) Euclidean space $\mathcal{R}^3$. By properly adopting a Cartesian coordinate system, we suppose the reference configuration of the plate occupies the region $\kappa_r=\Omega_r\times[0,h]$ in $\mathcal{R}^3$, where the thickness $h$ is much smaller than the dimensions of the in-plane area $\Omega_r$. The unit vector system directing along the coordinate axes is denoted as $\{\mathbf{e}_1,\mathbf{e}_2,\mathbf{k}\}$. For a given material point in the plate with coordinates $(X,Y,Z)$, its position vector is $\mathbf{R}=X\mathbf{e}_1+Y\mathbf{e}_2+Z\mathbf{k}$.

Suppose the plate undergoes biaxial growth along the $X$- and $Y$-axes. In this case, the growth field in the plate can be represented by tensor $\mathbb{G}=\mathrm{diag}(\lambda_1(X,Y,Z),\lambda_2(X,Y,Z),1)$, where $\lambda_1(X,Y,Z)$ and $\lambda_2(X,Y,Z)$ are the growth functions. In the current work, we assume that the growth field has the linear distribution form along the thickness direction of the plate, i.e.,
$$
\begin{aligned}
&\lambda_1(X,Y,Z)=\lambda_1^{(0)}(X,Y)+\lambda_1^{(1)}(X,Y)Z,\\
&\lambda_2(X,Y,Z)=\lambda_2^{(0)}(X,Y)+\lambda_2^{(1)}(X,Y)Z.\ \
\end{aligned}
\eqno(1)
$$
Despite the simplicity of these growth functions, it will be shown that they are able to yield arbitrary target shapes of the plate after growth.

As the growth field in the plate may be incompatible, it will induce residual stresses and further result in elastic deformations of the plate. After the deformation, the plate attains the current configuration $\kappa_t$. Suppose the material point in the plate has the new position vector $\mathbf{r}=x\mathbf{e}_1+y\mathbf{e}_2+z\mathbf{k}$ in $\kappa_t$, where the current coordinates $(x,y,z)$ depend on the referential coordinates $(X,Y,Z)$ in $\kappa_r$. The total deformation gradient tensor can then be calculated through
$$
\mathbb{F}=\frac{\partial\mathbf{r}}{\partial\mathbf{R}}=\nabla\mathbf{r}+\mathbf{r}_{,Z}\otimes\mathbf{k}=\mathbf{r}_{,X}\otimes\mathbf{e}_1+\mathbf{r}_{,Y}\otimes\mathbf{e}_2+\mathbf{r}_{,Z}\otimes\mathbf{k},\ \ \eqno(2)
$$
where `$\nabla$' is in-plane two-dimensional (2D) gradient operator. The subscripts `$,X$', `$,Y$' and `$,Z$' denote the derivatives along the coordinate axes. Following the approach proposed in Rodriguez \emph{et al.} \cite{rodr1994}, the total deformation gradient tensor can be decomposed into $\mathbb{F}=\mathbb{A}\mathbb{G}$, where $\mathbb{A}$ is the elastic strain tensor. As the elastic responses of soft materials are generally isochoric (e.g., soft biological tissues, polymeric gels), the elastic strain tensor should satisfy the following constraint equation
$$
R(\mathbb{F},\mathbb{G})=R_0(\mathbb{A})=\mathrm{Det}(\mathbb{A})-1=0.\ \ \eqno(3)
$$

We further assume that the plate is made of an incompressible neo-Hookean material. The elastic strain-energy function of the material is $\phi(\mathbb{F},\mathbb{G})=J_{G}\phi_0(\mathbb{A})=J_GC_0\left(\mathrm{tr}(\mathbb{A}\mathbb{A}^T)-3\right)$, where $J_{G}=\operatorname{Det}(\mathbb{G})$ and $C_{0}$ is a material constant. From the elastic strain-energy function, the nominal stress tensor $\mathbb{S}$ is given by
$$
\mathbb{S}=\frac{\partial\phi}{\partial\mathbb{F}}=J_{G}\mathbb{G}^{-1}\left(2C_{0}\mathbb{A}^{T}-p\mathbb{A}^{-1}\right),
\eqno(4)
$$
where $p(X,Y,Z)$ is the Lagrange multiplier associated with the constraint (3).

During the growth process, the hyperelastic plate satisfies the mechanical equilibrium equation
$$
\mathrm{Div}(\mathbb{S})=\mathbf{0},\ \ \mathrm{in}\ \ \kappa_r.\ \eqno(5)
$$
The faces of the plate are supposed to be traction-free, which yields the boundary conditions
$$
\begin{aligned}
&\mathbb{S}^T\mathbf{N}_l=\mathbf{0},\ \ \ \mathrm{on}\ \ \partial\Omega_r\times[0,h],\\
&\mathbb{S}^T\mathbf{k}|_{Z=0,h}=\mathbf{0},\ \ \ \mathrm{on}\ \ \Omega_r,
\end{aligned}
\eqno(6)
$$
where $\mathbf{N}_l$ is the unit normal vector on the lateral face of the plate. Eqs. (3), (5) and (6) formulate the 3D governing system of the plate model, which contains the unknowns $\mathbf{r}$ and $p$.


\subsection{Plate equation system}
\label{sec:2.2}

Starting from the 3D governing system, a consistent finite-strain plate equation system can be derived through a series expansion and truncation approach, which has been introduced in Wang et al. \cite{wang2018}. For being self-contained of the current paper, the key steps in the derivation procedure are listed below:
\begin{itemize}

\item Under the assumption of sufficient smoothness of $\mathbf{r}$ and $p$, we conduct series expansions of these unknowns along the thickness of the plate ($Z$-axis), i.e.,
$$
\begin{aligned}
&\mathbf{r}(X,Y,Z)=\sum_{n=0}^{2}\frac{Z^{n}}{n!}\mathbf{r}^{(n)}(X,Y)+O(Z^3),\ \\
&x(X,Y,Z)=\sum_{n=0}^{2}\frac{Z^{n}}{n!}x^{(n)}(X,Y)+O(Z^3),\ \\
&y(X,Y,Z)=\sum_{n=0}^{2}\frac{Z^{n}}{n!}y^{(n)}(X,Y)+O(Z^3),\ \\
&z(X,Y,Z)=\sum_{n=0}^{2}\frac{Z^{n}}{n!}z^{(n)}(X,Y)+O(Z^3),\ \\
&p(X,Y,Z)=\sum_{n=0}^{2}\frac{Z^n}{n!}p^{(n)}(X,Y)+O(Z^3),
\end{aligned}
\eqno(7)
$$
where $\mathbf{r}^{(n)}=x^{(n)}\mathbf{e}_1+y^{(n)}\mathbf{e}_2+z^{(n)}\mathbf{k}$ $(n=0,1,2)$. Corresponding to the expansions given in (7), the deformation gradient tensor $\mathbb{F}$, the elastic strain tensor $\mathbb{A}$ and the nominal stress tensor $\mathbb{S}$ can also be expanded as
$$
\begin{aligned}
&\mathbb{F}=\mathbb{F}^{(0)}+Z\mathbb{F}^{(1)}+O(Z^{2}),\\ \
&\mathbb{A}=\mathbb{A}^{(0)}+Z\mathbb{A}^{(1)}+O(Z^{2}),\\ \
&\mathbb{S}=\mathbb{S}^{(0)}+Z\mathbb{S}^{(1)}+O(Z^{2}),
\end{aligned}
\eqno(8)
$$
By using the kinematic relation (2), we obtain $\mathbb{F}^{(n)}=\nabla\mathbf{r}^{(n)}+\mathbf{r}^{(n+1)}\otimes\mathbf{k}$ $(n=0,1)$. Further from the relation $\mathbb{F}=\mathbb{A}\mathbb{G}$ and the constitutive relation (4), the explicit expressions of $\mathbb{A}^{(n)}$ and $\mathbb{S}^{(n)}$ $(n=0,1)$ can also be derived (cf. Eqs. (15) and (18) in Wang et al. \cite{wang2018}). Further from the mechanical equilibrium equation (5), we have the relation
$$
\nabla\cdot\mathbb{S}^{(n)}+\left(\mathbb{S}^{(n+1)}\right)^T\mathbf{k}=0,\
\eqno(9)
$$
In the current work, only the explicit expressions of $\mathbb{S}^{(0)}$ and $\mathbb{S}^{(1)}$ are required, which corresponds to $n=0$ in (9). In fact, if the expansion (8)$_3$ contains the high-order terms $\mathbb{S}^{(i)}$ $(i=2,3,\cdots)$, the relation (9) also holds for these high-order terms.

\item We substitute (7) into the constraint equation (3), the mechanical equilibrium equation (5) and the boundary condition (6)$_2$ at $Z=0$. By equating the coefficients of $Z^0$, $Z^1$ in (3) and $Z^0$ in (5) to be zero, combining with the boundary condition (6)$_2$, a closed linear system for the unknowns $\{\mathbf{r}^{(1)},\mathbf{r}^{(2)},p^{(0)},p^{(1)}\}$ is formulated. This linear system can be solved directly, then the following expressions of $\{\mathbf{r}^{(1)},\mathbf{r}^{(2)},p^{(0)},p^{(1)}\}$ in terms of $\mathbf{r}^{(0)}$ are obtained
$$
\begin{aligned}
&\mathbf{r}^{(1)}=\frac{\Lambda^{(0)}}{\Delta}\mathbf{N}, \ \ \ \ p^{(0)}=\frac{{\Lambda^{(0)}}^2}{\Delta},\\
&\mathbf{r}^{(2)}=-\frac{\mathbf{\bar{h}}}{\Lambda^{(0)}}+\left(\frac{\Lambda^{(1)}}{\Delta^2}-\frac{{\Lambda^{(0)}}^2\bar{\mathbf{s}}\cdot\mathbf{r}_N}{\Delta^6}+\frac{\mathbf{\bar{h}}\cdot\mathbf{r}_N}{\Lambda^{(0)}\Delta^2}\right)\mathbf{r}_N,\\
&p^{(1)}=2C_0\left(\frac{\Lambda^{(0)}\Lambda^{(1)}}{\Delta^2}-\frac{{\Lambda^{(0)}}^3\bar{\mathbf{s}}\cdot\mathbf{r}_N}{\Delta^6}+\frac{\mathbf{\bar{h}}\cdot\mathbf{r}_N}{\Delta^2}\right),
\end{aligned}
\eqno(10)
$$
where
$$
\begin{aligned}
&\Lambda^{(0)}=\lambda_1^{(0)}\lambda_2^{(0)},\ \ \ \Lambda^{(1)}=\lambda_1^{(1)}\lambda_2^{(0)}+\lambda_2^{(1)}\lambda_1^{(0)},\ \ \ \mathbf{r}_N=\mathbf{r}^{(0)}_{,X}\times\mathbf{r}^{(0)}_{,Y},\\
&\Delta=\sqrt{\mathbf{r}_N\cdot\mathbf{r}_N},\ \ \ \mathbf{N}=\frac{\mathbf{r}_N}{\Delta},\ \ \ \mathbf{s}_1=\mathbf{r}_N\times\mathbf{r}^{(0)}_{,X},\ \ \ \ \mathbf{s}_2=\mathbf{r}_N\times\mathbf{r}^{(0)}_{,Y},\\
&\bar{\mathbf{s}}=\mathbf{r}_{N,X}\times\mathbf{r}^{(0)}_{,Y}-\mathbf{r}_{N,Y}\times\mathbf{r}^{(0)}_{,X}, \ \ \ \mathbf{t}_1=\frac{\lambda_2^{(0)}}{\lambda_1^{(0)}}\mathbf{r}^{(0)}_{,X},\ \ \ \mathbf{t}_2=\frac{\lambda_1^{(0)}}{\lambda_2^{(0)}}\mathbf{r}^{(0)}_{,Y}, \\
&\mathbf{q}_1=\frac{{\Lambda^{(0)}}^2}{\Delta^2}\mathbf{r}_N\times\mathbf{r}^{(0)}_{,X},\ \ \ \mathbf{q}_2=\frac{{\Lambda^{(0)}}^2}{\Delta^2}\mathbf{r}_N\times\mathbf{r}^{(0)}_{,Y},\\
&\mathbf{\bar{h}}=\mathbf{t}_{1,X}+\mathbf{t}_{2,Y}-\frac{{\Lambda^{(0)}}^3}{\Delta^4}\bar{\mathbf{s}}+\frac{\Lambda^{(0)}\Lambda^{(1)}}{\Delta^2}\mathbf{r}_N+\frac{\Lambda^{(0)}}{\Delta^2}\left(\mathbf{q}_{2,X}-\mathbf{q}_{1,Y}\right).
\end{aligned}
\eqno(11)
$$

\item By using the relations given in (10), the stress tensors $\mathbb{S}^{(0)}$ and $\mathbb{S}^{(1)}$ in (8)$_3$ can be rewritten as
$$
\begin{aligned}
\mathbb{S}^{(0)}=&2C_0\left(\frac{{\Lambda^{(0)}}^3}{\Delta^4}\mathbf{r}_N\times\mathbf{r}^{(0)}_{,Y}+\frac{\lambda_2^{(0)}}{\lambda_1^{(0)}}\mathbf{r}^{(0)}_{,X}\right)\otimes\mathbf{e}_1\\
&+2C_0\left(-\frac{{\Lambda^{(0)}}^3}{\Delta^4}\mathbf{r}_N\times\mathbf{r}^{(0)}_{,X}+\frac{\lambda_1^{(0)}}{\lambda_2^{(0)}}\mathbf{r}^{(0)}_{,Y}\right)\otimes\mathbf{e}_2,
\end{aligned}
\eqno(12)
$$
$$
\begin{aligned}
\mathbb{S}^{(1)}=2C_0\Bigg[&\frac{{\Lambda^{(0)}}^4}{\Delta^6}\mathbf{r}_N\times\mathbf{r}_{N,Y}+\frac{\Lambda^{(0)}}{\Delta^2}\mathbf{r}^{(0)}_{,Y}\times\mathbf{\bar{h}}+\frac{\lambda_2^{(1)}\lambda_1^{(0)}-\lambda_1^{(1)}\lambda_2^{(0)}}{{\lambda_1^{(0)}}^2}\mathbf{r}^{(0)}_{,X}\\
&+\frac{2\Lambda^{(0)}}{\Delta^4}\left(\Lambda^{(0)}\Lambda^{(1)}-\frac{{\Lambda^{(0)}}^3\bar{\mathbf{s}}\cdot\mathbf{r}_N}{\Delta^4}+\mathbf{\bar{h}}\cdot\mathbf{r}_N\right)\mathbf{r}_N\times\mathbf{r}^{(0)}_{,Y}\\
&+\frac{\lambda_2^{(0)}}{\lambda_1^{(0)}}\frac{\partial}{\partial X}\left(\frac{\Lambda^{(0)}}{\Delta^2}\mathbf{r}_N\right)\Bigg]\otimes\mathbf{e}_1\\
+2C_0\Bigg[&-\frac{{\Lambda^{(0)}}^4}{\Delta^6}\mathbf{r}_N\times\mathbf{r}_{N,X}-\frac{\Lambda^{(0)}}{\Delta^2}\mathbf{r}^{(0)}_{,X}\times\mathbf{\bar{h}}+\frac{\lambda_1^{(1)}\lambda_2^{(0)}-\lambda_2^{(1)}\lambda_1^{(0)}}{{\lambda_2^{(0)}}^2}\mathbf{r}^{(0)}_{,X}\\
&-\frac{2\Lambda^{(0)}}{\Delta^4}\left(\Lambda^{(0)}\Lambda^{(1)}-\frac{{\Lambda^{(0)}}^3\bar{\mathbf{s}}\cdot\mathbf{r}_N}{\Delta^4}+\mathbf{\bar{h}}\cdot\mathbf{r}_N\right)\mathbf{r}_N\times\mathbf{r}^{(0)}_{,X}\\
&+\frac{\lambda_1^{(0)}}{\lambda_2^{(0)}}\frac{\partial}{\partial Y}\left(\frac{\Lambda^{(0)}}{\Delta^2}\mathbf{r}_N\right)\Bigg]\otimes\mathbf{e}_2\\
+2C_0\Bigg[&-\mathbf{\bar{h}}+\frac{\Lambda^{(0)}\Lambda^{(1)}}{\Delta^2}\mathbf{r}_N+\frac{{\Lambda^{(0)}}^2}{\Delta^2}\bigg[-\frac{\partial}{\partial X}\left(\frac{\Lambda^{(0)}}{\Delta^2}\mathbf{r}_N\times\mathbf{r}^{(0)}_{,Y}\right)\\
&+\frac{\partial}{\partial Y}\left(\frac{\Lambda^{(0)}}{\Delta^2}\mathbf{r}_N\times\mathbf{r}^{(0)}_{,X}\right)\bigg]\Bigg]\otimes\mathbf{e}_3,
\end{aligned}
\eqno(13)
$$

\item Subtracting the top and bottom boundary conditions given in (6)$_2$, then by virtue of the relation (9), the following vectorial plate equation can be established
$$
\nabla\cdot\overline{\mathbb{S}}=0, \quad \mathrm{in}\ \Omega_r,
\eqno(14)
$$
where
$$
\overline{\mathbb{S}}=\frac{1}{h}\int_{0}^{h}\mathbb{S}dZ=\mathbb{S}^{(0)}+\frac{h}{2}\mathbb{S}^{(1)}+O\left(h^{2}\right). \eqno(15)
$$
By substituting (12) and (13) into (14), we obtain three plate equations for the three components of $\mathbf{r}^{(0)}$ (i.e., $\{x^{(0)},y^{(0)},z^{(0)}\}$). To complete the plate equation system, we propose the following boundary conditions on the edge of the in-plane area $\partial\Omega_r$
$$
\begin{aligned}
&\overline{\mathbb{S}}^{T}\mathbf{N}_l=\mathbf{0},\\
&\mathbf{M}_h=\frac{1}{h}\int_{0}^{h}\left(\mathbb{S}^{T}\mathbf{N}_l\right)\times\left[\mathbf{r}-\mathbf{r}|_{Z=h/2}\right]dZ=\mathbf{0},
\end{aligned}
\eqno(16)
$$
where $\mathbf{M}_h$ is the bending moment about the middle plane $Z=h/2$ of the plate.

\end{itemize}


\section{Shape-programming of thin hyperelastic plates}
\label{sec:3}


\subsection{Growth functions in the stress-free condition}
\label{sec:3.1}

The plate equation system has been established in the previous section. For any given growth functions $\lambda_1^{(n)}$ and $\lambda_2^{(n)}$ $(n=0,1)$, one can solve this plate equation system, then the growth-induced deformations of the thin hyperelastic plate will be predicted. In the current work, we aim to solve an inverse problem. That is, to achieve certain target configuration of the thin hyperelastic plate through differential growth, how to arrange the growth fields in the plate? This problem is referred to as `shape-programming' of thin hyperelastic plates \cite{liu2016}.

\begin{figure}
  \centering \includegraphics[width=0.8\textwidth]{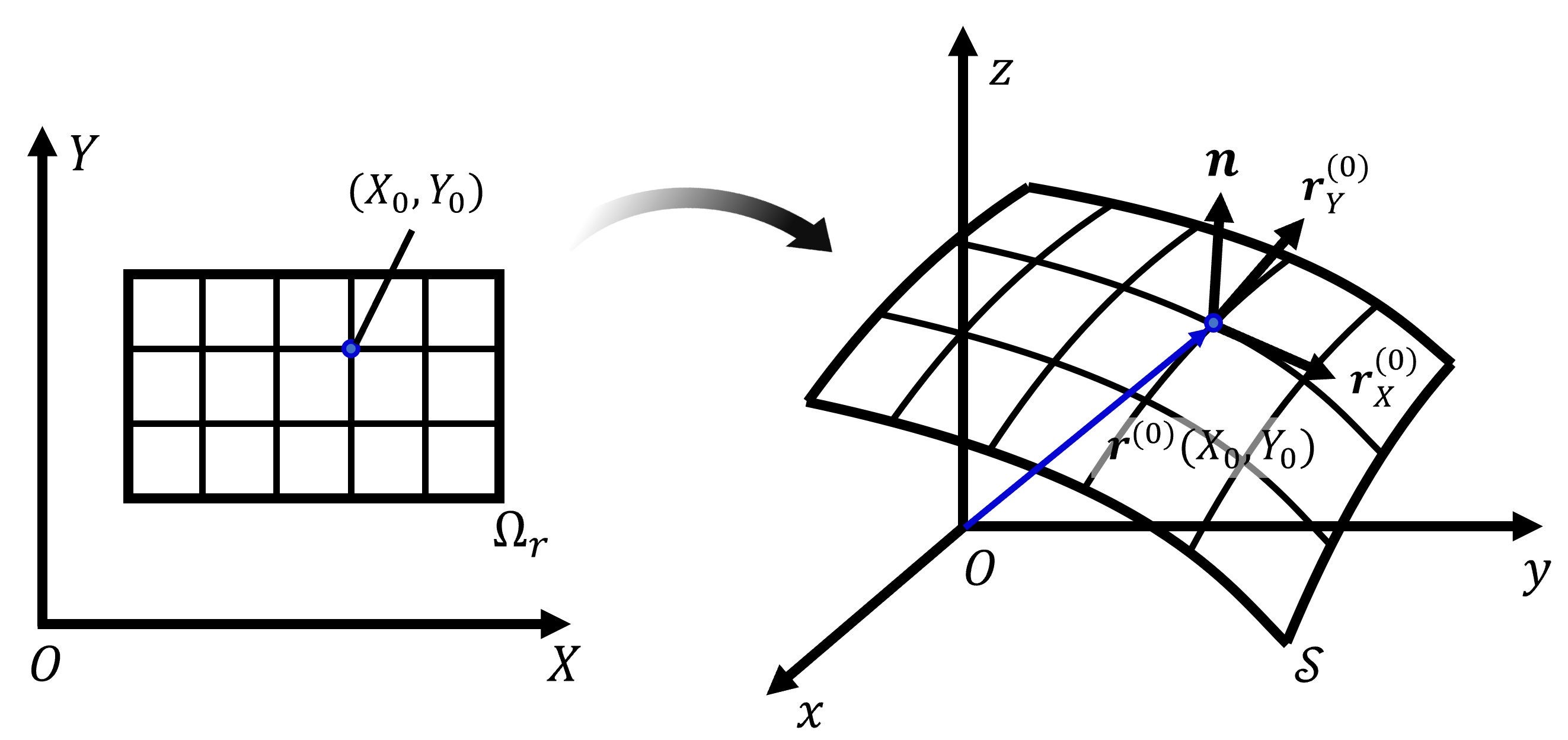}
  \caption{Illustration of the mapping $\mathbf{r}^{(0)}$ from the in-plane area $\Omega_r$ to the surface $\mathcal{S}$.}
\label{fig:1}
\end{figure}

It should be pointed out that we do not aim to control the whole 3D configuration of the plate. As the plate equation system (14)-(16) is derived based on the bottom face ($Z=0$) of the plate, shape-programming will also be conducted by only taking the bottom face into account. In the current configuration $\kappa_t$, the original flat bottom face $\Omega_r$ has transformed into a surface $\mathcal{S}\subset\mathcal{R}^3$, which has the following parametric equation
$$
\mathbf{r}^{(0)}(X,Y)=(x^{(0)}(X,Y),y^{(0)}(X,Y),z^{(0)}(X,Y)), \ \ \ (X,Y) \in \Omega_r. \ \ \eqno(17)
$$
Eq. (17) can be viewed as a continuous mapping from $\Omega_r$ to $\mathcal{S}$ (cf. Fig. \ref{fig:1}). For convenience of the following analyses, we assume that the functions $x^{(0)}(X,Y)$, $y^{(0)}(X,Y)$ and $z^{(0)}(X,Y)$ have sufficient smoothness. With the given value of one variable $X_0$ or $Y_0$, $\mathbf{r}^{(0)}(X,Y_0)$ and $\mathbf{r}^{(0)}(X_0,Y)$ generate the so called $X$-curve and $Y$-curve on the surface accompanying the variation of the other variable. All of these curves formulate the parametric curves net on $\mathcal{S}$. At any point $\mathbf{r}^{(0)}(X_0,Y_0)$, the tangent vectors along the $X$- and $Y$-curves can be represented by $\mathbf{r}^{(0)}_{,X}|_{(X_0,Y_0)}$ and $\mathbf{r}^{(0)}_{,Y}|_{(X_0,Y_0)}$. We further assume that $\mathbf{r}^{(0)}_{,X}\times\mathbf{r}^{(0)}_{,Y}\neq0$ at any point on $\mathcal{S}$, which implies that $\mathcal{S}$ is a regular surface.

To achieve the goal of shape-programming, one needs to determine the growth functions $\lambda_1^{(n)}$ and $\lambda_2^{(n)}$ $(n=0,1)$ corresponding to any target surface $\mathcal{S}$, such that the plate equation system is satisfied. Generally, the solution of shape-programming may not be unique \cite{wang2019}. In this section, we focus on the case that the hyperelastic plate has the stress-free state in $\kappa_t$, i.e., all the components in $\mathbb{S}^{(0)}$ and $\mathbb{S}^{(1)}$ are vanished. It's clear that in the stress-free condition, both the plate equations (14) and the boundary conditions (16) are automatically satisfied (some recent studies on growth-induced deformations of hyperelastic bodies with stress-free states can be found in Chen and Dai \cite{chen2020}). Next, we begin to study the relations between the growth functions and the geometrical properties of the target surface $\mathcal{S}$.

First, we consider the case that all the components in $\mathbb{S}^{(0)}$ are vanished. From (12), we have the following two vectorial equations
$$
\begin{aligned}
\frac{{\Lambda^{(0)}}^3}{\Delta^4}\mathbf{r}_N\times\mathbf{r}^{(0)}_{,Y}+\frac{\lambda_2^{(0)}}{\lambda_1^{(0)}}\mathbf{r}^{(0)}_{,X}=\mathbf{0},\\
-\frac{{\Lambda^{(0)}}^3}{\Delta^4}\mathbf{r}_N\times\mathbf{r}^{(0)}_{,X}+\frac{\lambda_1^{(0)}}{\lambda_2^{(0)}}\mathbf{r}^{(0)}_{,Y}=\mathbf{0}.
\end{aligned}
\eqno(18)
$$
By conducting the dot products of (18)$_1$ with $\mathbf{r}^{(0)}_{,X}$ and (18)$_2$ with $\mathbf{r}^{(0)}_{,Y}$, it can be obtained that
$$
\begin{aligned}
&\mathbf{r}^{(0)}_{,X}\cdot\mathbf{r}^{(0)}_{,X}=-\frac{{\Lambda^{(0)}}^3\lambda_1^{(0)}\left[\left(\mathbf{r}_N\times\mathbf{r}^{(0)}_{,Y}\right)\cdot\mathbf{r}^{(0)}_{,X}\right]}{\lambda_2^{(0)}\Delta^4}=\frac{{\lambda_1^{(0)}}^4{\lambda_2^{(0)}}^2}{\mathbf{r}_N\cdot\mathbf{r}_N},\\
&\mathbf{r}^{(0)}_{,Y}\cdot\mathbf{r}^{(0)}_{,Y}=\frac{{\Lambda^{(0)}}^3\lambda_2^{(0)}\left[\left(\mathbf{r}_N\times\mathbf{r}^{(0)}_{,X}\right)\cdot\mathbf{r}^{(0)}_{,Y}\right]}{\lambda_1^{(0)}\Delta^4}=\frac{{\lambda_1^{(0)}}^2{\lambda_2^{(0)}}^4}{\mathbf{r}_N\cdot\mathbf{r}_N},
\end{aligned}
\eqno(19)
$$
Besides that, we also have equality
$$
\begin{aligned}
\mathbf{r}_N\cdot\mathbf{r}_N&=\left(\mathbf{r}^{(0)}_{,X}\times\mathbf{r}^{(0)}_{,Y}\right)\cdot\left(\mathbf{r}^{(0)}_{,X}\times\mathbf{r}^{(0)}_{,Y}\right)\\
&=\left(\mathbf{r}^{(0)}_{,X}\cdot\mathbf{r}^{(0)}_{,X}\right)\left(\mathbf{r}^{(0)}_{,Y}\cdot\mathbf{r}^{(0)}_{,Y}\right)-\left(\mathbf{r}^{(0)}_{,X}\cdot\mathbf{r}^{(0)}_{,Y}\right)^2=EG-F^2.
\end{aligned}
\eqno(20)
$$
Here, we denote $E=\mathbf{r}^{(0)}_{,X}\cdot\mathbf{r}^{(0)}_{,X}$, $F=\mathbf{r}^{(0)}_{,X}\cdot\mathbf{r}^{(0)}_{,Y}$ and $G=\mathbf{r}^{(0)}_{,Y}\cdot\mathbf{r}^{(0)}_{,Y}$. It is known that these three quantities are just the \emph{coefficients of the first fundamental form} of the surface $\mathcal{S}$. By substituting (20) into (19) and solving the two equations, we obtain
$$
\lambda_1^{(0)}=\frac{E^{\frac{1}{3}}(EG-F^2)^{\frac{1}{6}}}{G^{\frac{1}{6}}},\ \ \ \lambda_2^{(0)}=\frac{G^{\frac{1}{3}}(EG-F^2)^{\frac{1}{6}}}{E^{\frac{1}{6}}}. \ \eqno(21)
$$
Especially, if the $X$- and $Y$-curves formulate the orthogonal parametric curves net on $\mathcal{S}$ (i.e., $F=\mathbf{r}^{(0)}_{,X}\cdot\mathbf{r}^{(0)}_{,Y}=0$), the growth functions can be simplified into
$$
\lambda_1^{(0)}=\sqrt{E},\ \ \ \ \lambda_2^{(0)}=\sqrt{G}. \ \eqno(22)
$$
Therefore, the growth functions $\lambda_1^{(0)}$ and $\lambda_2^{(0)}$ just represent the in-plane extension or shrinkage of the plate during the transformation from $\Omega_r$ to $\mathcal{S}$.

Second, we consider the case that all the components in $\mathbb{S}^{(1)}$ are vanished. From the expression of $\mathbb{S}^{(1)}$ given in (13), three vectorial equations are obtained. Here, we still adopt the assumption of orthogonality of the parametric curves on $\mathcal{S}$. By using (20) and (22), we have $\Lambda^{(0)}=\sqrt{EG}=\Delta$. Therefore, the three equations are simplified into
$$
\begin{aligned}
&\frac{\mathbf{r}_N\times\mathbf{r}_{N,Y}}{\Delta^2}+\frac{\mathbf{r}^{(0)}_{,Y}\times\mathbf{\bar{h}}}{\Delta}+\frac{\lambda_2^{(1)}\lambda_1^{(0)}-\lambda_1^{(1)}\lambda_2^{(0)}}{{\lambda_1^{(0)}}^2}\mathbf{r}^{(0)}_{,X}\\
&+\left(\frac{2\Lambda^{(1)}}{\Delta^2}-\frac{2\bar{\mathbf{s}}\cdot\mathbf{r}_N}{\Delta^4}+\frac{2\mathbf{\bar{h}}\cdot\mathbf{r}_N}{\Delta^3}\right)\mathbf{r}_N\times\mathbf{r}^{(0)}_{,Y}+\frac{\lambda_2^{(0)}}{\lambda_1^{(0)}}\frac{\partial}{\partial X}\left(\frac{\mathbf{r}_N}{\Delta}\right)=\mathbf{0},\\
&-\frac{\mathbf{r}_N\times\mathbf{r}_{N,X}}{\Delta^2}-\frac{\mathbf{r}^{(0)}_{,X}\times\mathbf{\bar{h}}}{\Delta}+\frac{\lambda_1^{(1)}\lambda_2^{(0)}-\lambda_2^{(1)}\lambda_1^{(0)}}{{\lambda_2^{(0)}}^2}\mathbf{r}^{(0)}_{,X}\\
&-\left(\frac{2\Lambda^{(1)}}{\Delta^2}-\frac{2\bar{\mathbf{s}}\cdot\mathbf{r}_N}{\Delta^4}+\frac{2\mathbf{\bar{h}}\cdot\mathbf{r}_N}{\Delta^3}\right)\mathbf{r}_N\times\mathbf{r}^{(0)}_{,X}+\frac{\lambda_1^{(0)}}{\lambda_2^{(0)}}\frac{\partial}{\partial Y}\left(\frac{\mathbf{r}_N}{\Delta}\right)=\mathbf{0},\\
&\mathbf{\bar{h}}-\frac{\Lambda^{(1)}\mathbf{r}_N}{\Delta}+\frac{\partial}{\partial X}\left(\frac{\mathbf{r}_N\times\mathbf{r}^{(0)}_{,Y}}{\Delta}\right)-\frac{\partial}{\partial Y}\left(\frac{\mathbf{r}_N\times\mathbf{r}^{(0)}_{,X}}{\Delta}\right)=\mathbf{0},
\end{aligned}
\eqno(23)
$$
By using the expressions of $\mathbf{r}_N$, $\bar{\mathbf{s}}$, $\mathbf{\bar{h}}$ and $\Lambda^{(1)}$ given in (11), it can be directly verified that (23)$_3$ is automatically satisfied, where the relations
$$
\mathbf{r}_N\times\mathbf{r}^{(0)}_{,X}={\lambda_1^{(0)}}^2\mathbf{r}^{(0)}_{,Y}, \ \ \ \mathbf{r}_N\times\mathbf{r}^{(0)}_{,Y}=-{\lambda_2^{(0)}}^2\mathbf{r}^{(0)}_{,X}
$$
are utilized. The other two equations in (23) can be rewritten as
$$
\begin{aligned}
&\Bigg[3\Lambda^{(1)}+\frac{2\left({\lambda_1^{(0)}}^2L+{\lambda_2^{(0)}}^2N\right)}{\lambda_1^{(0)}\lambda_2^{(0)}}+\frac{2\lambda_2^{(0)}L}{\lambda_1^{(0)}}-(\lambda_2^{(1)}\lambda_1^{(0)}-\lambda_1^{(1)}\lambda_2^{(0)})\Bigg]\mathbf{r}^{(0)}_{,X}\\
&=-\frac{2\lambda_1^{(0)}M}{\lambda_2^{(0)}}\mathbf{r}^{(0)}_{,Y},
\end{aligned}
\eqno(24)
$$
$$
\begin{aligned}
&\Bigg[3\Lambda^{(1)}+\frac{2\left({\lambda_1^{(0)}}^2L+{\lambda_2^{(0)}}^2N\right)}{\lambda_1^{(0)}\lambda_2^{(0)}}+\frac{2\lambda_1^{(0)}N}{\lambda_2^{(0)}}+(\lambda_2^{(1)}\lambda_1^{(0)}-\lambda_1^{(1)}\lambda_2^{(0)})\Bigg]\mathbf{r}^{(0)}_{,Y}\\
&=-\frac{2\lambda_2^{(0)}M}{\lambda_1^{(0)}}\mathbf{r}^{(0)}_{,X},
\end{aligned}
\eqno(25)
$$
where we denote $L=\mathbf{r}^{(0)}_{,XX}\cdot\mathbf{N}$, $M=\mathbf{r}^{(0)}_{,XY}\cdot\mathbf{N}$ and $N=\mathbf{r}^{(0)}_{,YY}\cdot\mathbf{N}$. It is known that $L$, $M$ and $N$ are just the \emph{coefficients of the second fundamental form} of the surface $\mathcal{S}$. To ensure the holds of Eqs. (24) and (25), we need to require $M=0$, which implies that the $X$- and $Y$-curves formulate the orthogonal curvature curves net on $\mathcal{S}$. Further from (24) and (25), it can be derived that
$$
\lambda_1^{(1)}=-\frac{L}{\lambda_1^{(0)}}, \ \ \ \ \lambda_2^{(1)}=-\frac{N}{\lambda_2^{(0)}}. \ \eqno(26)
$$
From (26), it can be seen that the growth functions $\lambda_1^{(1)}$ and $\lambda_2^{(1)}$ are closely related to the curvatures of the target surface $\mathcal{S}$.


\subsection{A theoretical scheme for shape-programming}
\label{sec:3.2}

Eqs. (22) and (26) provide the relations between the growth functions and the coefficients of first and second fundamental forms of surface $\mathcal{S}$. It is known that the surface can be uniquely identified (up to a rigid body motion) by the first and second fundamental forms \cite{chen2017,topo2006}. Thus, corresponding to an arbitrary target surface $\mathcal{S}$, the growth fields in the plate just need to be arranged according to (22) and (26). However, these relations are derived based on the assumption that the parametric coordinate curves formulate an orthogonal curvature curves net. Usually, this requirement is not satisfied by the given parametric equation $\mathbf{r}^{(0)}(X,Y)$. In this case, some manipulations should be conducted in advance to generate the orthogonal curvature curves net on the surface $\mathcal{S}$.

\begin{figure}
  \centering \includegraphics[width=0.8\textwidth]{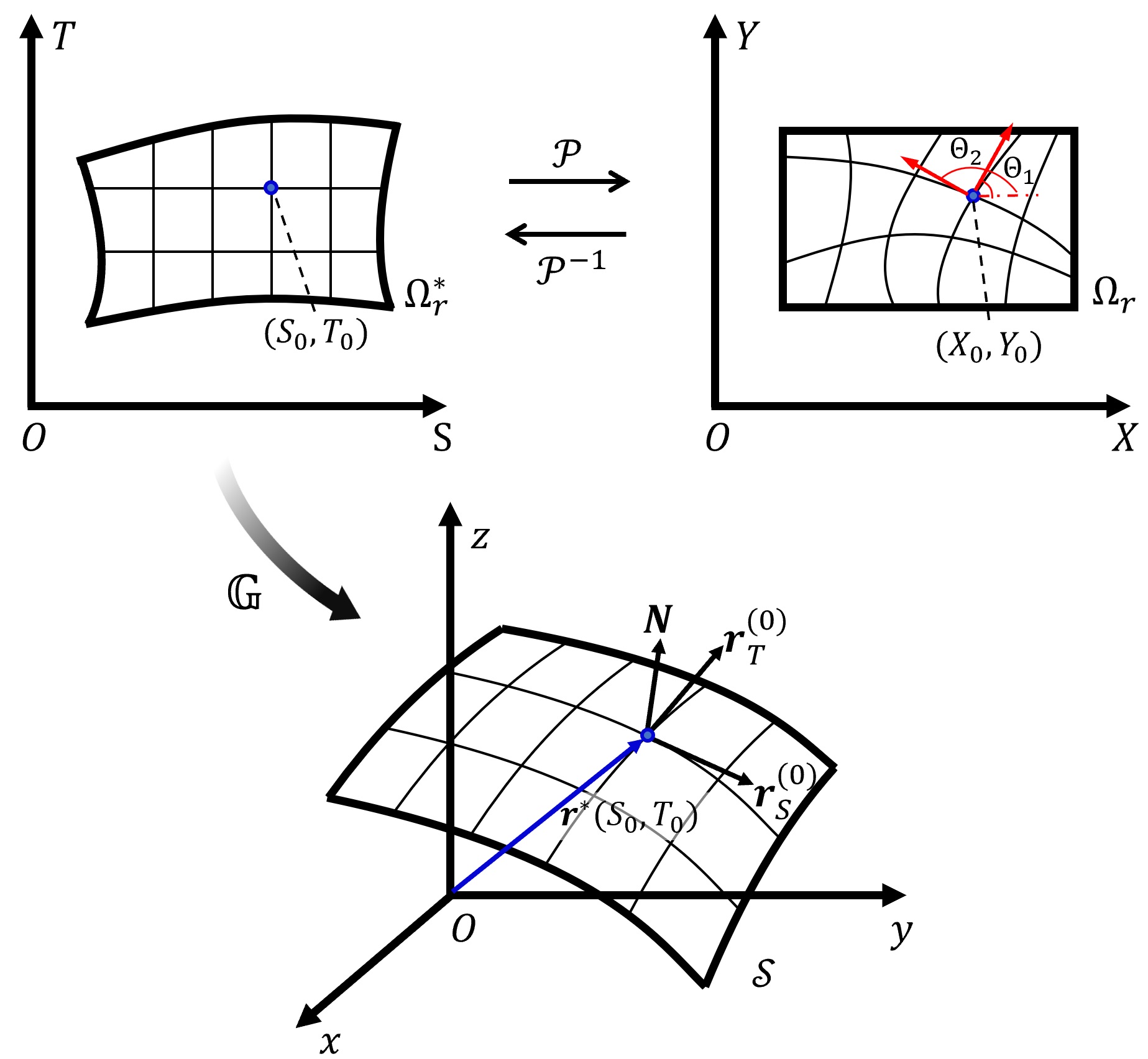}
  \caption{Illustration of the coordinate transformation between $\{X,Y\}$ and $\{S,T\}$, and the mapping $\mathbf{r}^{*}$ from the in-plane area $\Omega_r^{*}$ to the surface $\mathcal{S}$.}
\label{fig:2}
\end{figure}

Consider a target surface $\mathcal{S}$ defined on $\Omega_r$, which has the parametric equation $\mathbf{r}^{(0)}(X,Y)$. If the $X$- and $Y$-curves cannot formulate an orthogonal curvature curves net on $\mathcal{S}$, we conduct the following variable transformation
$$
X=X(S,T),\ \ \ Y=Y(S,T),\ \ \eqno(27)
$$
where $X(S,T)$ and $Y(S,T)$ are sufficient smooth and the Jacobi determinant $\partial(X,Y)/\partial(S,T)>0$. In fact, based on the transformation (27), a bijection between $\Omega_r$ in the $XY$-plane and a new region $\Omega_r^{*}$ in the $ST$-plane (cf. Fig. \ref{fig:2}). Through the variable transformation, $\mathcal{S}$ has a new parametric equation $\mathbf{r}^{*}(S,T)=\mathbf{r}^{(0)}(X(S,T),Y(S,T))$, from which we have
$$
\begin{aligned}
&\mathbf{r}_{,S}^{*}=\mathbf{r}^{(0)}_{,X}\frac{\partial X}{\partial S}+\mathbf{r}^{(0)}_{,Y}\frac{\partial Y}{\partial S}=A_1\left(\mathbf{r}^{(0)}_{,X}\cos\Theta_1+\mathbf{r}^{(0)}_{,Y}\sin\Theta_1\right),\\
&\mathbf{r}_{,T}^{*}=\mathbf{r}^{(0)}_{,X}\frac{\partial X}{\partial T}+\mathbf{r}^{(0)}_{,Y}\frac{\partial Y}{\partial T}=A_2\left(\mathbf{r}^{(0)}_{,X}\cos\Theta_2+\mathbf{r}^{(0)}_{,Y}\sin\Theta_2\right),\\
\end{aligned}
\eqno(28)
$$
where
$$
\begin{aligned}
A_1=\sqrt{\left(\frac{\partial X}{\partial S}\right)^2+\left(\frac{\partial Y}{\partial S}\right)^2},\ \ \ \cos\Theta_1=\frac{\frac{\partial X}{\partial S}}{A_1}, \ \ \ \sin\Theta_1=\frac{\frac{\partial Y}{\partial S}}{A_1},\\
A_2=\sqrt{\left(\frac{\partial X}{\partial T}\right)^2+\left(\frac{\partial Y}{\partial T}\right)^2},\ \ \ \cos\Theta_2=\frac{\frac{\partial X}{\partial T}}{A_2}, \ \ \ \sin\Theta_2=\frac{\frac{\partial Y}{\partial T}}{A_2},\\
\end{aligned}
\eqno(29)
$$
To ensure the parametric coordinate curves (i.e., $S$- and $T$-curves) generated from $\mathbf{r}^{*}(S,T)$ formulate an orthogonal curvature curves net, $\mathbf{r}_{,S}^{*}$ and $\mathbf{r}_{,T}^{*}$ should be aligned with the principle directions at any point $\mathbf{r}^{*}(S_0,T_0)$ on $\mathcal{S}$. Correspondingly, $\Theta_1$ and $\Theta_2$ defined in (29) satisfy the following equation \cite{chen2017,topo2006}
$$
(LF-ME){\cos^2\Theta}+(LG-NE)\cos\Theta\sin\Theta+(MG-NF){\sin^2\Theta}=0,\ \ \eqno(30)
$$
where $\{E,F,G\}$ and $\{L,M,N\}$ are the coefficients of fundamental forms calculated from the original parametric equation $\mathbf{r}^{(0)}(X,Y)$. On the other hand, as the transformation between $\{X,Y\}$ and $\{S,T\}$ is a bijection, we have
$$
\begin{aligned}
\left(\begin{array}{cc}
\frac{\partial S}{\partial X} & \frac{\partial S}{\partial Y}\\
\frac{\partial T}{\partial X} & \frac{\partial T}{\partial Y}\\
\end{array}
\right)&=\left(\begin{array}{cc}
\frac{\partial X}{\partial S} & \frac{\partial X}{\partial T}\\
\frac{\partial Y}{\partial S} & \frac{\partial Y}{\partial T}\\
\end{array}
\right)^{-1}\\
&=\left(\begin{array}{cc}
A_1^*\sin\Theta_2 & -A_1^*\cos\Theta_2\\
-A_2^*\sin\Theta_1 & A_2^*\cos\Theta_1\\
\end{array}
\right),
\end{aligned}
\eqno(31)
$$
where
$$
\begin{aligned}
&A_1^*=\frac{1}{A_1(\cos\Theta_1\sin\Theta_2-\sin\Theta_1\cos\Theta_2)}, \\
&A_2^*=\frac{1}{A_1(\cos\Theta_1\sin\Theta_2-\sin\Theta_1\cos\Theta_2)}.
\end{aligned}
$$
Next, we consider the following differential forms
$$
\begin{aligned}
&dS=\frac{\partial S}{\partial X}dX+\frac{\partial S}{\partial Y}dY=A_1^*\left(\sin\Theta_2dX-\cos\Theta_2dY\right),\\
&dT=\frac{\partial T}{\partial X}dX+\frac{\partial T}{\partial Y}dY=A_2^*\left(-\sin\Theta_1dX+\cos\Theta_1dY\right).
\end{aligned}
\eqno(32)
$$
To obtain the explicit expressions of the transformation between $\{X,Y\}$ and $\{S,T\}$, one needs to find the integrating factors $A_1^*$ and $A_2^*$ such that the differential forms $dS$ and $dT$ given in (32) are integrable. Then, the first integrals of these differential forms just provides the explicit expressions of $S(X,Y)$ and $T(X,Y)$. Accordingly, the expressions of $X=X(S,T)$ and $Y=Y(S,T)$ are also obtained. If the functions $\{\sin\Theta_i,\cos\Theta_i\}_{i=1,2}$ are continuously differentiable and they are not both equal to zero at certain point $(X_0,Y_0)\in\Omega_r$, it has been proved that the integrating factor $A_i^*$ must exist in a neighboring region of $(X_0,Y_0)$ \cite{chen2017}. However, to our knowledge, there is still no universal formulas to provide the integrating factors for any differential forms. In some specific cases, the integrating factors can be derived by adopting suitable techniques.

\begin{figure}
  \centering \includegraphics[width=0.55\textwidth]{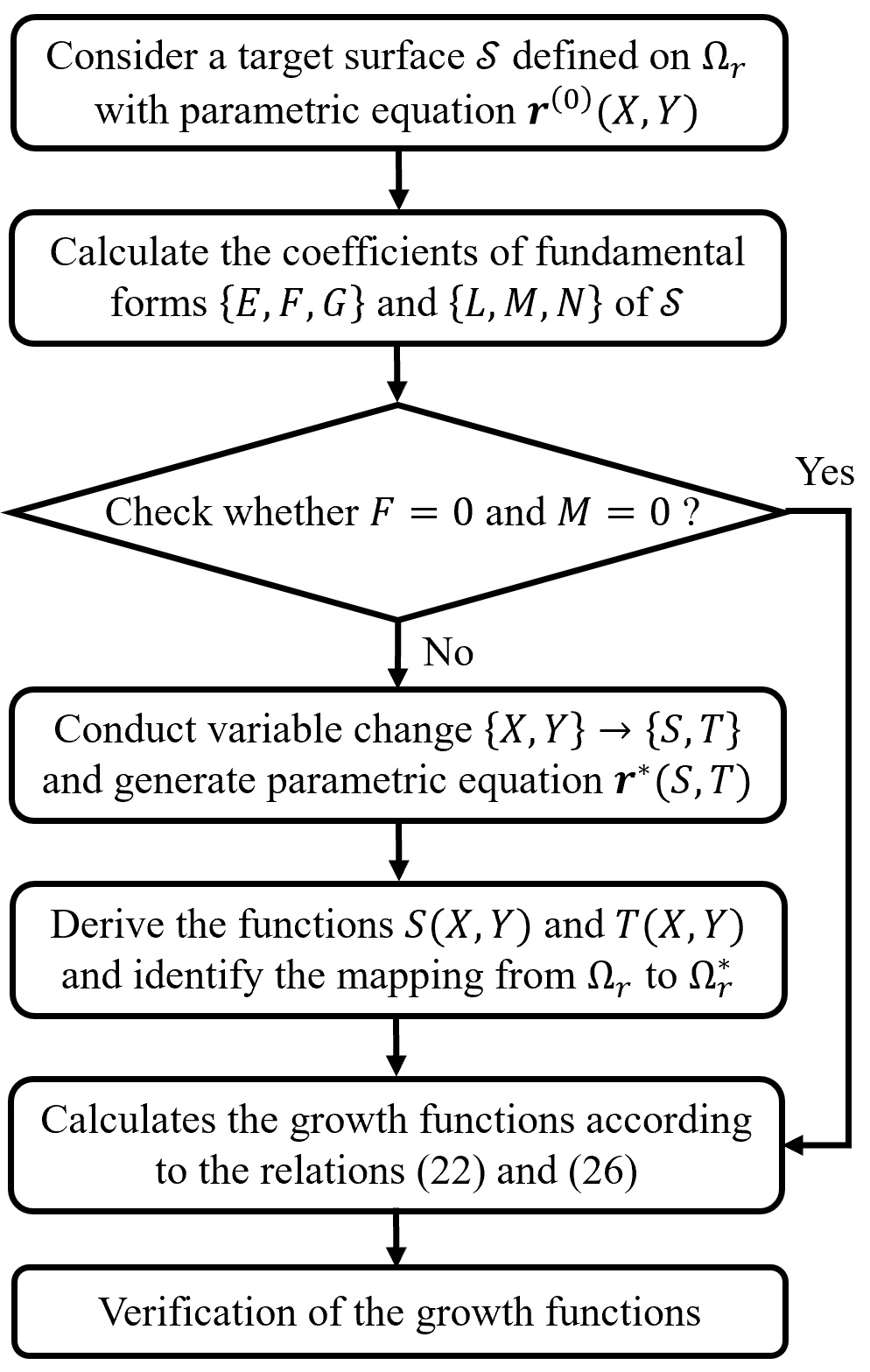}
  \caption{Flowchart of the scheme of shape-programming of a thin hyperelastic plate through differential growth.}
\label{fig:3}
\end{figure}

Based on the above preparations, we can propose a theoretical scheme for shape-programming of a thin hyperelastic plate through differential growth. The flowchart of this scheme is shown in Fig. \ref{fig:3}. First, we consider a target surface $\mathcal{S}$ with the parametric equation $\mathbf{r}^{(0)}(X,Y)$, which is defined on the in-plane area $\Omega_r$. To check whether the parametric coordinate curves obtained from $\mathbf{r}^{(0)}(X,Y)$ formulate an orthogonal curvature curves net on $\mathcal{S}$, we calculate the coefficients $\{E,F,G\}$ and $\{L,M,N\}$ of the first and second fundamental forms of $\mathcal{S}$. In the case $F=0$ and $M=0$, it is known that the parametric curves net is already an orthogonal curvature curves net \cite{chen2017}. Thus, the relations (22) and (26) can be directly used to calculate the growth functions $\lambda_1^{(n)}$ and $\lambda_2^{(n)}$ $(n=0,1)$. If $F$ and $M$ are not both equal to zero, we need to conduct the variable transformation from $\{X,Y\}$ to $\{S,T\}$ and generate a new parametric equation $\mathbf{r}^{*}(S,T)$, which yields a bijective mapping from $\Omega_r$ to a new region $\Omega_r^*$ in the $ST$-plane. To ensure that the parametric coordinate curves of $\mathbf{r}^{*}(S,T)$ formulate an orthogonal curvature curves net on $\mathcal{S}$, the functions $\Theta_1$ and $\Theta_2$ should be determined from the equation (30). After that, we need to find proper integrating factors $A_1^*$ and $A_2^*$ for the differential forms given in (32), based on which the explicit expressions of $S(X,Y)$ and $T(X,Y)$ can be derived. With the new parametric equation $\mathbf{r}^{*}(S,T)$, the growth functions can also be calculated from the relations (22) and (26). Finally, to check the correctness and efficiency of this scheme, the obtained growth functions will be incorporated in a finite element program and the growth-induced deformations of the plate will be simulated.

\noindent\textbf{Remark}: In the theoretical scheme proposed in the section, we always assume the target surface $\mathcal{S}$ has an initial parametric equation $\mathbf{r}^{(0)}(X,Y)$. However, the surfaces with complex geometrical shapes usually have no explicit parametric equations. In that case, some numerical schemes need to be designed to determine the distributions of growth fields in the thin hyperelastic plates.


\section{Application examples}
\label{sec:4}

To demonstrate the efficiency of the theoretical scheme of shape-programming, some typical examples will be studied in this section. In these examples, for any given target surface $\mathcal{S}$ with the initial parametric equation $\mathbf{r}^{(0)}(X,Y)$, the growth functions in the plate are calculated according to the proposed theoretical scheme, which are further adopted in the numerical simulations to verify their correctness.

\begin{figure}
  \centering \includegraphics[width=0.70\textwidth]{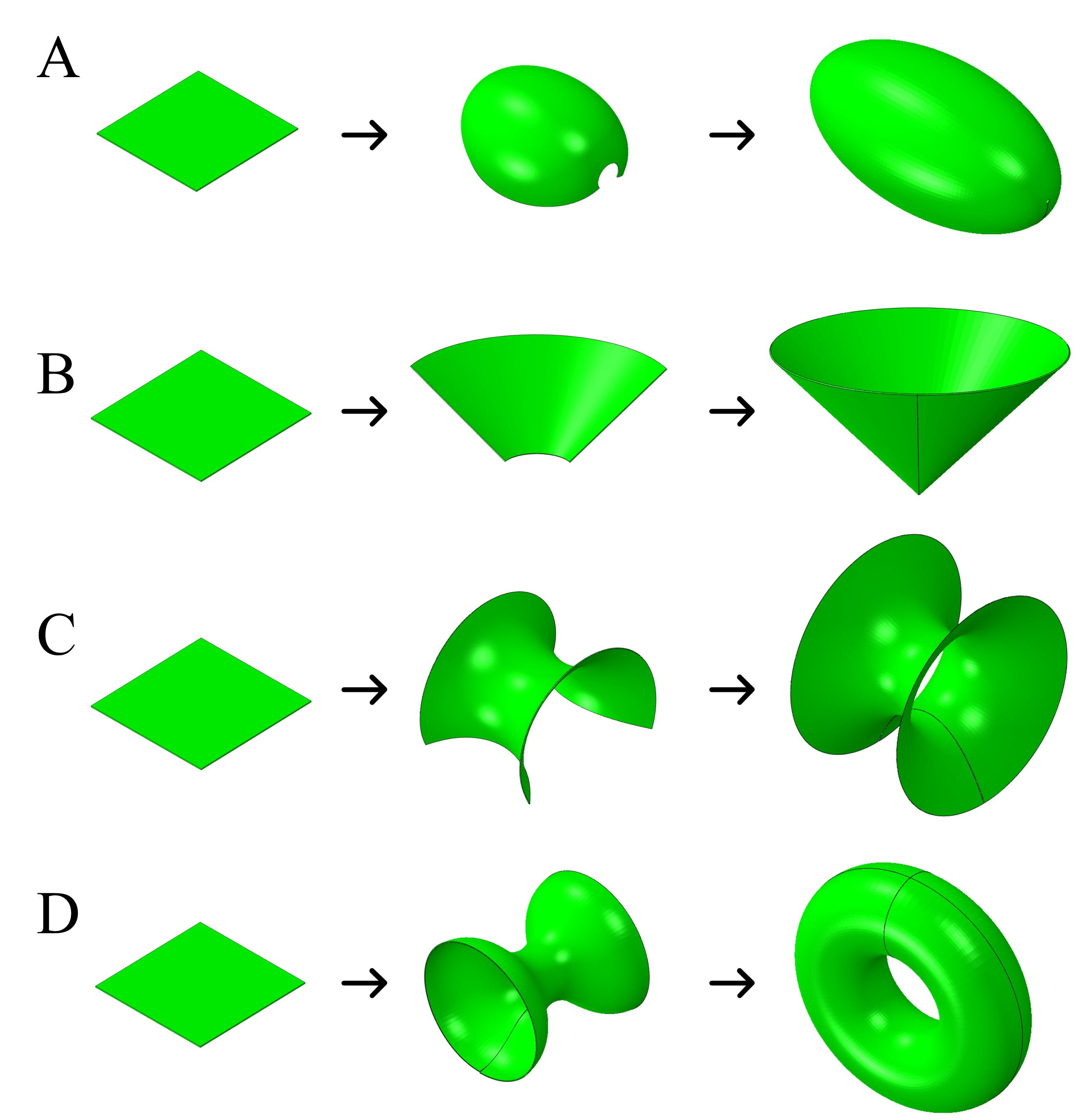}
  \caption{Numerical simulations results on the growing processes of the plate: (A) the ellipsoid surface; (B) the conical surface; (C) the catenoid surface; (D) the toroidal surface.}
\label{fig:4}
\end{figure}

In the first example, we select the rotating surface as the target surface $\mathcal{S}$, which has the following parametric equation
$$
\mathbf{r}^{(0)}(X,Y)=(f(X)\cos(2\pi Y),f(X)\sin(2\pi Y),g(X)),\ \ \eqno(33)
$$
where $f(X)$ and $g(X)$ are some arbitrary smooth functions. Corresponding to this parametric equation, the coefficients of first and second fundamental forms can be calculated, which are given by
$$
\begin{aligned}
&E={f_{,X}}^2+{g_{,X}}^2,\ \ \ \ F=0,\ \ \ \ G=4\pi^2{f}^2,\\
&L=\frac{f(f_{,X}g_{,XX}-g_{,X}f_{,XX})}{\sqrt{f^2({f_{,X}}^2+{g_{,X}}^2)}},\ \ \ \ M=0,\ \ \ \ N=\frac{4\pi^2f^2g_{,X}}{\sqrt{f^2({f_{,X}}^2+{g_{,X}}^2)}}.
\end{aligned}
\eqno(34)
$$
It can be seen that the conditions $F=0$ and $M=0$ have already been satisfied. Thus, the parametric coordinate curves generated from the parametric equation (33) can formulate the orthogonal curvature curves net on the surface. From the relations (22) and (26), we obtain the following growth functions
$$
\begin{aligned}
&\lambda_1^{(0)}=\sqrt{{f_{,X}}^2+{g_{,X}}^2},\ \ \ \lambda_2^{(0)}=2\pi|f|,\\
&\lambda_1^{(1)}=-\frac{f_{,X}g_{,XX}-g_{,X}f_{,XX}}{|f|\left({f_{,X}}^2+{g_{,X}}^2\right)},\ \ \ \lambda_2^{(1)}=-\frac{2\pi g_{,X}}{{f_{,X}}^2+{g_{,X}}^2}.\\
\end{aligned}
\eqno(35)
$$
For the purpose of illustration, we consider four kinds of rotating surfaces, i.e., the ellipsoid surface, the conical surface, the catenoid surface and the toroidal surface. The parametric equations and the corresponding growth functions of these surfaces are listed in (36), where the initial in-plane region $\Omega_r$ is chosen to be $\Omega_r=[0,1]\times[0,1]$. To verify the correctness of the obtained growth functions, we further conduct numerical simulations by using the FEM software ABAQUS. A modified compressible neo-Hookean material model is incorporated in the UMAT subroutine of ABAQUS, which contains the growth functions ${\lambda_1(X,Y,Z),\lambda_2(X,Y,Z)}$ as the state variables. During the numerical calculations, UMAT subroutine is called at each integration point of the elements. With the input data of displacements and state variables, the total deformation gradient tensor $\mathbb{F}$ and the growth tensor $\mathbb{G}$ can be determined, then the elastic strain tensor $\mathbb{A}$ is calculated from $\mathbb{A}=\mathbb{F}\mathbb{G}^{-1}$. With the obtained elastic strain tensor, the Cauchy stress tensor, the consistent Jacobian are updated, which are output to the FE program for further calculations. To simulate the whole growing process, the growth functions $\lambda_1(X,Y,Z)$ and $\lambda_2(X,Y,Z)$ changes linearly from $1$ to the specified values. The material constants in the model are chosen such that the Poisson's ratio $\mu=0.4995$ (i.e., close to the incompressibility condition). The reference configuration of the plate is set to be $[0,1]\times[0,1]\times[0,0.01]$, which is meshed into $20000$ C3D8IH (an 8-node linear brick, hybrid, linear pressure, incompatible modes) elements. To capture the out-of-plane deformations of the plate, certain buckling mode multiplied by a damping factor is applied to the plate as initial geometric imperfection. In Fig. \ref{fig:4}, we show the numerical simulation results on the growth-induced deformations of the plate. It can be seen that in these four cases, the grown states of the plate can fit the target surfaces quit well, thus the correctness of the obtained growth functions can be verified.

\begin{itemize}

\item Ellipsoid surface ($0\leq X\leq1, 0\leq Y\leq1$)
$$
\left\{
\begin{aligned}
&x^{(0)}=\sin(\pi X)\cos(2\pi Y),\ \ \ y^{(0)}=\sin(\pi X)\sin(2\pi Y),\ \ \\
&z^{(0)} = 2\cos(\pi X),\\
&\lambda_1 = \frac{\pi}{\sqrt{2}}\sqrt{5-3\cos(2 \pi X)}+\frac{4\pi Z}{5-3\cos(2\pi X)},\\
&\lambda_2 = 2\pi\sin(\pi X) + \frac{4\sqrt{2}\pi\sin(\pi X)Z}{\sqrt{5-3\cos(2\pi X)}},\\
\end{aligned}\right.
\eqno(36)_1
$$

\item Conical surface ($0\leq X\leq1, 0\leq Y\leq1$)
$$
\left\{
\begin{aligned}
&x^{(0)}=X\sin(2\pi Y),\ \ \ y^{(0)}=X\cos(2\pi Y),\ \ \ z^{(0)}=X,\\
&\lambda_1 = \sqrt{2},\ \ \ \lambda_2 = 2\pi X+\sqrt{2}\pi Z,
\end{aligned}\right.
\eqno(36)_2
$$

\item Catenoid surface ($0\leq X\leq1, 0\leq Y\leq1$)
$$
\left\{
\begin{aligned}
&x^{(0)} = -2\cosh\left(\pi X-\frac{\pi}{2}\right)\cos(2\pi Y),\\
&y^{(0)} = -2\cosh\left(\pi X-\frac{\pi}{2}\right)\sin(2\pi Y),\\
&z^{(0)} = \pi(2X-1),\\
&\lambda_1 = \sqrt{2}\pi\sqrt{\cosh(\pi-2\pi X)+1} - \pi Z \mathrm{sech}\left(\frac{\pi}{2}-\pi X\right),\\
&\lambda_2 = 2\sqrt{2}\pi\sqrt{\cosh(\pi -2 \pi X)+1} + 2\pi Z\mathrm{sech}\left(\frac{\pi}{2}-\pi X\right),
\end{aligned}\right.
\eqno(36)_3
$$

\item Toroidal surface ($0\leq X\leq1, 0\leq Y\leq1$)
$$
\left\{
\begin{aligned}
&x^{(0)} = \frac{1}{2}[\cos(2\pi X)+2]\cos(2\pi Y),\\
&y^{(0)} = \frac{1}{2}[\cos(2\pi X)+2]\sin(2\pi X),\\
&z^{(0)} = \frac{1}{2}\sin(2\pi X),\\
&\lambda_1 = \pi + 2\pi Z,\\
&\lambda_2 = \pi[2+\cos(2\pi X)] + 2\pi\cos(2\pi X)Z,
\end{aligned}\right.
\eqno(36)_4
$$
\end{itemize}

\begin{figure}
  \centering \includegraphics[width=0.75\textwidth]{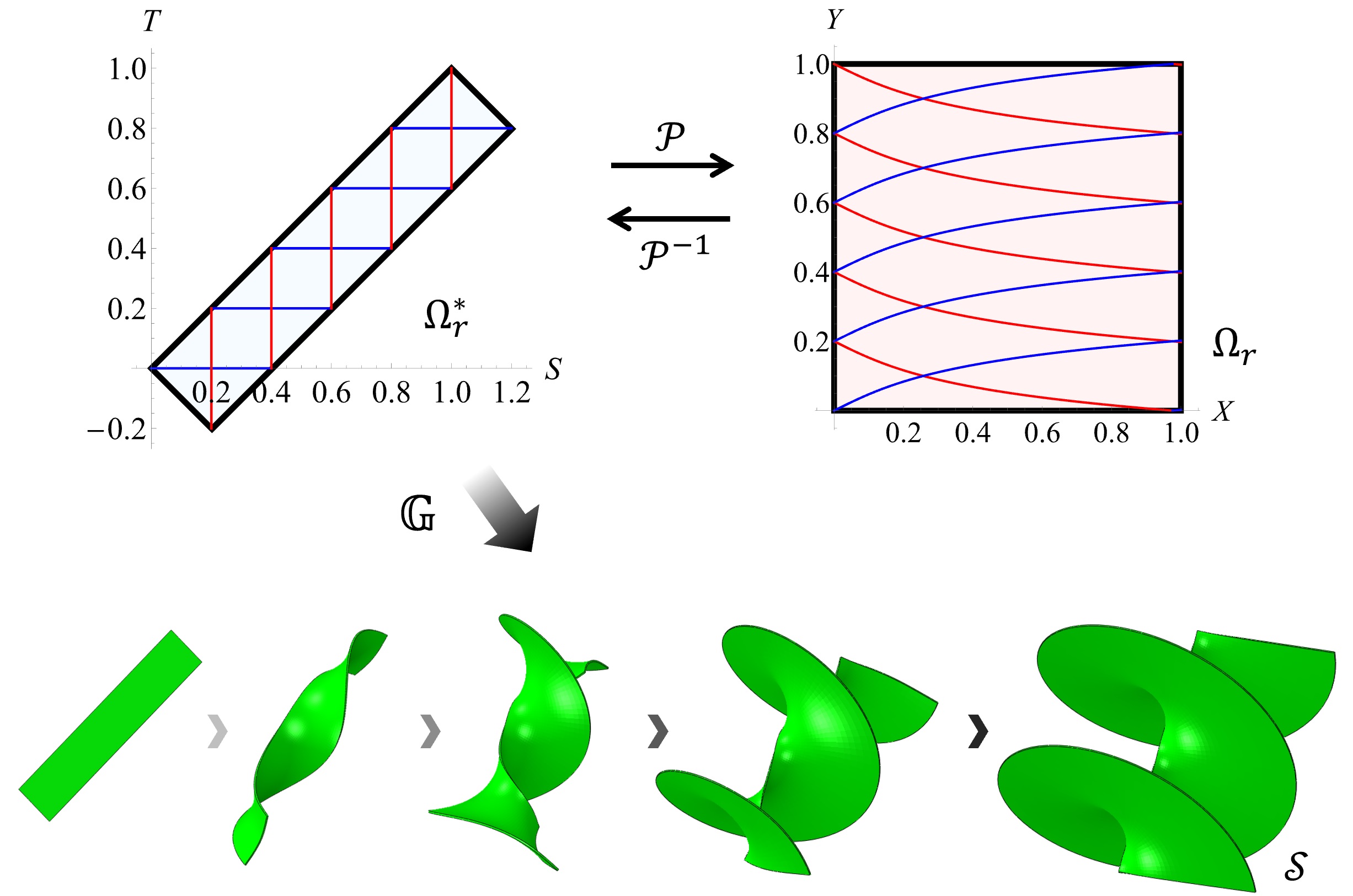}
  \caption{The variable transformation between $\{X,Y\}$ and $\{S,T\}$ and the numerical simulation of the growing process to generate the helical configuration of the plate.}
\label{fig:5}
\end{figure}

In the second example, we select a helical surface as the target surface $\mathcal{S}$, which has the following parametric equation
$$
\mathbf{r}^{(0)}(X,Y)=(X\sin(4\pi Y),X\cos(4\pi Y),2Y),\ \ \ \eqno(37)
$$
where the initial in-plane region $\Omega_r$ is still chosen to be $\Omega_r=[0,1]\times[0,1]$. From the parametric equation (37), it is found that the coefficient of the second fundamental form $M=4\pi/\sqrt{1+4\pi^2X^2}\neq0$, thus the relations (22) and (26) cannot be used directly. We need to conduct the variable transformation from $\{X,Y\}$ to $\{S,T\}$. According to the scheme introduced in the previous section, it can be derived that
$$
\begin{aligned}
&\sin\Theta_1=\frac{1}{\sqrt{5+16\pi^2X^2}},\ \ \ \ \cos\Theta_1=\frac{\sqrt{4+16\pi^2X^2}}{\sqrt{5+16\pi^2X^2}},\\
&\sin\Theta_2=-\frac{1}{\sqrt{5+16\pi^2X^2}},\ \ \ \ \cos\Theta_2=\frac{\sqrt{4+16\pi^2X^2}}{\sqrt{5+16\pi^2X^2}},
\end{aligned}
\eqno(38)
$$
Then, the integrating factors of the differential forms (32) can be chosen as
$$
A_1^{*}=-\sqrt{1+\frac{1}{4+16\pi^2X^2}},\ \ \ \ A_2^{*}=\sqrt{1+\frac{1}{4+16\pi^2X^2}}. \eqno(39)
$$
By substituting (39) into (32), we obtain the following explicit expressions of variable transformation
$$
S(X,Y)=\frac{\mathrm{arcsinh}(2\pi X)}{4\pi}+Y,\ \ \ \ T(X,Y)=-\frac{\mathrm{arcsinh}(2\pi X)}{4\pi}+Y.\ \ \eqno(40)
$$
In this transformation, the original in-plane region $\Omega_r$ is mapped into a new region $\Omega_r^{*}$ in the $ST$-plane, which are shown in Fig. \ref{fig:5}. By using (37) and (40), the new parametric equation of the helical surface based on the variables $\{S,T\}$ can be obtained. Based on this new parametric equation, the growth functions can be calculated from (22) and (26), which are given by
$$
\begin{aligned}
\lambda_1=&\sqrt{1+\cosh(4\pi(S-T))}\\
&-2\pi Z \sqrt{[1+\cosh(4\pi(S-T))]\mathrm{sech}^4(2\pi(S-T))},\\
\lambda_2=&\sqrt{1+\cosh(4\pi(S-T))}\\
&+2\pi Z \sqrt{[1+\cosh(4\pi(S-T))]\mathrm{sech}^4(2\pi(S-T))}.
\end{aligned}
\eqno(41)
$$
To verify the correctness of these growth functions, we also conduct numerical simulations on the growing process of the thin hyperelastic plate. The setting of numerical calculation is same as that introduced in the first example, only except that the reference configuration of the hyperelastic plate is chosen to be $\Omega_r^{*}\times[0,0.01]$. The simulated grown state of the plate is shown in Fig. \ref{fig:5}, which can also fit the target surface quite well.

\section{Conclusions}

In this paper, the deformations of thin hyperelastic plates induce by differential growth were investigated. To achieve the goal of shape-programming of hyperelastic plates, we proposed a theoretical scheme to determine the growth functions corresponding to any 3D target surfaces. The following tasks have been accomplished: (1) a consistent finite-strain plate equation system for growth-induced deformations of a neo-Hookean plate sample was formulated; (2) under the stress-free condition, the inverse problem was solved analytically, from which the relations between growth functions and geometrical properties (i.e., the first and second fundamental forms) of the target surface were revealed; (3) a theoretical for shape-programming of thin hyperelastic plates through differential growth was proposed; (4) the correctness and efficiency of the scheme was verified through some typical examples. Since the obtained explicit formulas for shape-programming have relatively simple forms, it will be useful for design and manufacture of intelligent soft devices. Furthermore, the analytical results can provide significant insight into the growth behaviors of some soft biological tissues in nature.

Besides the above advantages, it should be pointed out that the analytical formulas for shape-programming were derived under the stress-free condition, which may not be applicable in the case that the plate is subjected to external loads or boundary restrictions. For some complicated surfaces that have no explicit parametric equations, the proposed theoretical scheme is also not applicable. To fulfill the requirements of practical applications, the problems with more general boundary conditions need to be investigated. In addition, a numerical scheme for shape-programming of complicated surfaces needs to be developed in the future.

\section*{Acknowledgments}


%
%
%

\bibliographystyle{unsrt}  
\bibliography{MMS}


\end{document}